\documentclass[prl,twocolumn,showpacs,superscriptaddress,amsfonts,amsmath,floatfix]{revtex4}
\usepackage{graphicx}
\newcommand{\Journal}[4]{#1 {\bf #2}, #3 (#4)}
\newcommand{\PRL}{Phys. Rev. Lett.}

\newcommand{\Science}{Science}

\begin{document}

\title{Three-body exclusion principle, duality mapping, and exact ground state of a harmonically trapped, ultracold Bose gas
with three-body hard-core interactions in one dimension}
\author{M. D. Girardeau}
\email{girardeau@optics.arizona.edu}
\affiliation{Kavli Institute for Theoretical Physics, University of California, Santa Barbara, CA 93106, USA}
\affiliation{College of Optical Sciences, University of Arizona, Tucson, AZ 85721, USA \footnote{Current and permanent address}}
\date{\today}
\begin{abstract}
Motivated by previous suggestions that three-body hard-core interactions in lower-dimensional ultracold Bose gases might provide a
way for creation of non-Abelian anyons, the exact ground state of a harmonically trapped 1D Bose gas with three-body hard-core
interactions is constructed by duality mapping, starting from an $N$-particle ideal gas of mixed symmetry with three-body nodes, which has double occupation of the lowest harmonic oscillator orbital and single occupation of the next $N-2$ orbitals. It has some similarity
to the ground state of a Tonks-Girardeau gas, but is more complicated. It is proved that in 1D any system of
$N\ge 3$ bosons with three-body hard-core interactions also has two-body soft-core interactions of generalized Lieb-Liniger
delta function form, as a consequence of the topology of the configuration space of $N$ particles in 1D, i.e., wave functions
with \emph{only} three-body hard core zeroes are topologically impossible. This is in contrast with the case in 2D, where pure 
three-body hard-core interactions do exist, and are closely related to the fractional quantized Hall effect. The exact ground state
is compared with a previously-proposed Pfaffian-like approximate ground state, which satisfies the three-body hard-core constraint
but is not an exact energy eigenstate. Both the exact ground state and the Pfaffian-like approximation imply two-body soft-core
interactions as well as three-body hard-core interactions, in accord with the general topological proof. 
\end{abstract}
\maketitle
If an ultracold atomic vapor is confined in a de Broglie wave guide with
transverse trapping so tight and temperature so low that the transverse
vibrational excitation quantum is larger than available
longitudinal zero point and thermal energies, the effective dynamics becomes
one-dimensional (1D) \cite{Ols98,PetShlWal00}. 3D Feshbach resonances \cite{Rob01}
allow tuning to the neighborhood of
1D confinement-induced resonances \cite{Ols98,BerMooOls03} where the 1D interaction is very
strong, leading to strong short-range correlations, breakdown of
effective-field theories, and emergence of highly-correlated $N$-body
ground states. In the case of spinless or spin-polarized bosons with 1D zero-range Lieb-Liniger (LL) \cite{LieLin63} delta function 
repulsion $g_2\delta(x_j-x_\ell)$ with coupling constant $g_2\to +\infty$, the
Tonks-Girardeau (TG) gas, the exact $N$-body ground state was determined in 1960 by a
Fermi-Bose (FB) mapping to an ideal Fermi gas \cite{Gir60}, leading to
``fermionization'' of many properties of this Bose system, as recently
confirmed experimentally \cite{Par04,Kin04}. 

Under conditions of ultracold gas experiments, the 1D two-body interaction Hamiltonian is usually well-approximated by the zero-range LL
potential $V_2=g_2\sum_{1\le j<k\le N}\delta(x_j-x_k)$, and the dimensionless coupling constant measuring its strength is
$\gamma_2=mg_2/n\hbar^2$ where $n$ is the 1D number density. $\gamma_2\ll 1$ is the Gross-Pitaevskii regime and $\gamma_2\gg 1$ is the
TG regime. A three-body zero-range interaction potential generalizing the LL-interaction is 
$V_3=g_3\sum_{1\le j<k<\ell\le N}\delta(x_j-x_k)\delta(x_j-x_\ell)$, and its dimensionless coupling constant is
$\gamma_3=mg_3/\hbar^2$. Note that $\gamma_3$ is independent of the density, whereas $\gamma_2$ has the density in the denominator.
Hence, attainment of the two-body TG limit $\gamma_2\gg 1$ requires very low density or very large $g_2$, the reason why attainment
of this limit is so difficult and the experiments \cite{Par04,Kin04} were \textit{tours de force}. This suggests that attainment
of the three-body hard-core limit $\gamma_3\gg 1$ may be easier than reaching the TG limit. Although three-body collisions are rare at the usual densities of ultracold gases, the magnitude of the three-body scattering length does not depend on density.

Laughlin's wave function  \cite{Lau83} for the ground state of a 2D electron gas in a magnetic field in the lowest Landau level is
$\exp(-\frac{1}{4\ell^2}\sum_{j=1}^N|z_j^2|)\prod_{1\le j<k\le N}(z_j-z_k)^\nu$ where $\nu$ is an odd integer. It is an ansatz describing
quasiparticles of fractional charge $e/\nu$ in the fractional quantum Hall effect. Here $z_j=x_j+iy_j$ are complex position coordinates in
the $(x,y)$ plane, $\ell^2=\hbar/m\omega_c$, and $\omega_c$ is the cyclotron frequency. In the limit of infinitely strong interaction
it is known \cite{Gre_etal,Rei96} that the \emph{exact} ground state for even $N$ is a closely related Pfaffian-like wave function
$\Psi_{0}=\hat{S}_{\uparrow,\downarrow}\exp(-\frac{1}{4\ell^2}\sum_{j=1}^N|z_j^2|)
\prod_{j<k}^{N/2}(z_j^\uparrow-z_k^\uparrow)^2\prod_{j<k}^{N/2}(z_j^\downarrow-z_k^\downarrow)^2$ where $\hat{S}_{\uparrow,\downarrow}$
symmetrizes over all ways of subdividing the $N$ particles into two subsets of $N/2$ each. It describes particles with 
2D three-body hard-core interactions, since  $\Psi_{0}$ vanishes when $z_j=z_k=z_\ell$ for all choices $j<k<\ell$. The strong similarity
between the Laughlin wave function and the 1D TG ground state 
$\psi_{B0}(x_{1},\cdots,x_{N})=\left[\prod_{i=1}^{N}e^{-Q_{i}^{2}/2}\right]\prod_{1\le j<k\le N}|x_{j}-x_{k}|$ 
in a harmonic trap \cite{GirWriTri01}, where  $Q_{i}=x_{i}/x_\text{osc}$ with $x_\text{osc}=\sqrt{\hbar/m\omega}$ the oscillator length,
suggests a possible close connection with the problem of three-body hard-core interactions in 1D, motivating Paredes \textit{et al.}
\cite{ParKeiCir07} to suggest a similar Pfaffian-like state as an ansatz for the 1D ground state. It was 
previously suggested \cite{Gre_etal,MooRea91} that the fractional quantum Hall effect in 2D might lead to a class of quasiparticles obeying non-Abelian anyon statistics, with potentially important applications to creation of topologically
protected qubits for quantum computation, and very recently experimental evidence for such quasiparticles has been found
\cite{WilPfeWes10}. Although no way is currently known for producing such quasiparticles in 1D, this is further motivation for
exploring connections between three-body interactions in 2D and those in 1D, as suggested in \cite{ParKeiCir07}. 

Paredes \textit{et al.} \cite{ParKeiCir07} assume
that $N$ is even, and construct an approximate ground state by dividing the $N$ particles into two subsets of $\frac{N}{2}$ each, 
assuming that each of these two is in an $\frac{N}{2}$-particle TG ground state, taking the product of these two states,
finally restoring the required bosonic symmetry over all $N!$ permutations by symmetrizing over all ways of choosing two subsets
of $\frac{N}{2}$ particles from all $N$. They assumed that the $N$ bosons were trapped on a ring, thus 
requiring the ring-periodic TG ground state \cite{Gir60}. However, the TG ground state in a harmonic trap is much simpler
\cite{GirWriTri01}. With that choice the ansatz of \cite{ParKeiCir07} is
$\Psi_{0\text{par}}=[\exp(-\sum_{j=1}^N m\omega x_j^2/2\hbar)]\hat{S}[\prod_{i<j}^{N/2}|x_i-x_j|]
[\prod_{k<\ell}^{N/2}|x_k-x_\ell|]$ where $\hat{S}$ is the symmetrizer described above. The similarity with the above 2D Pfaffian-like
state is evident. This state vanishes when $x_j=x_k=x_\ell$ for all $1\le j<k<\ell\le N$, because any such choice requires that
at least two of these three coordinates, say $x_p$ and $x_q$, lie in the same subset, giving a vanishing factor $|x_p-x_q|$ with 
$x_p=x_q$. However, it is not an energy eigenstate. Numerical
calculations described in \cite{ParKeiCir07} suggest that its error does not exceed 10\%. In the remainder of this paper the 
\emph{exact} ground state with three-body hard-core interactions will be constructed by combining a three-body exclusion principle 
with a duality mapping generalizing that of \cite{Gir60}.

\textit{Three-body exclusion principle and ideal gas with three-body nodes:} The key to finding the exact ground state of a 1D
Bose gas with three-body hard-core interactions is a three-body exclusion principle and its application to generation of 
three-body nodes in an ideal gas of mixed symmetry. The Pauli exclusion principle requires that all $N$-particle wave functions of
a system of fermions must be totally antisymmetric under all permutations of its particle coordinates $X_j$, where $X_j$ includes
both the spatial position and any discrete internal quantum numbers. For a system of particles without internal quantum numbers,
one consequence is that all allowed wave functions vanish if two or more particles have the same spatial position. Although
a corollary of this is that allowed wave functions vanish if three particles are at the same point, this is an utterly 
trivial consequence. A much weaker requirement is that allowed wave functions must vanish if \emph{three or more} particles
are at the same point, with no restriction on the wave function if two are at the same point. This defines a "three-body exclusion
principle", and is the starting point for the determination herein of the exact ground state of bosons with three-body hard-core
interactions in 1D. The Hamiltonian $\hat{H}_0$ of $N$ identical particles in 1D with no interparticle interactions in a harmonic trap consists of only the kinetic energy and the trap potential:
\begin{equation}\label{free}
\hat{H}_0=\sum_{j=1}^N\left(-\frac{\hbar^2}{2m}\frac{\partial^2}{\partial x_j^2}+\frac{1}{2}m\omega^2 x_j^2\right)\ .
\end{equation}
Its eigenstates are trapped ideal gas states, and the first goal here is to find ideal gas states of mixed symmetry satisfying the
three-body exclusion principle. The single-particle energy eigenstates in the trap are the orbitals 
$\phi_n(x)=c_n\exp(-m\omega^2 x^2/2\hbar)H_n(x/x_\text{osc})$ where $H_n$ are the Hermite polynomials and $c_n$ is a 
normalization constant. Start from an unsymmetrized orbital product state $\Psi_{0U}$ with two particles in the lowest orbital and
one in each of the remaining $N-2$ lowest orbitals:
\begin{equation}\label{unsymm}
\Psi_{0U}=\phi_0(x_1)\phi_0(x_2)\phi_1(x_3)\cdots\phi_{N-2}(x_N)\ .
\end{equation}
It is an eigenstate of $\hat{H}_0$ with energy
\begin{equation}\label{ground energy}
E_0=(\frac{1}{2}+\frac{1}{2}+\frac{3}{2}+\cdots+\frac{2N-3}{2})\hbar\omega=\frac{1}{2}(N^2-2N+2)\hbar\omega\ ,
\end{equation}
and this will remain true if its $N$ arguments $(x_1,\cdots,x_N)$ are permuted in any way, and for all linear combinations of such
products. Our goal is to choose linear combinations summing to a mixed-symmetry "model state" $\Psi_{0M}$ which satisfies the
three-body exclusion principle. This requirement is satisfied by choosing coefficients $\pm 1$ in such a way that $\Psi_{0M}$ is
the sum of $N$ terms, each of which is the product of the lowest orbital $\phi_0(x_j)$, with $j$ ranging from $1$ to $N$, and a
Slater determinant with the remaining $N-1$ atoms occupying each of $\phi_0,\phi_1,\cdots,\phi_{N-2}$ once:
\begin{equation}
\Psi_{0M}(x_1,\cdots,x_N)=\sum_{j=1}^N\phi_0(x_j)\text{det}_{(n,k=0,1),k\ne j}^{(N-1,N)}\phi_n(x_k)\ .
\end{equation}
The antisymmetry of each Slater determinant ensures that if more than two particles are at the same point $x_j=x_k=x_\ell=x$,
$\Psi_{0M}$ will vanish, i.e. this $(N-3)$-dimensional hyperline is a three-body node of $\Psi_{0M}$. If $x_j=x_k=x$ and $x_\ell$
is in the neighborhood of $x$, then $\Psi_{0M}$ changes sign as $x_\ell$ passes through $x$, i.e., it is locally but not globally
antisymmetric about that point. Next, note that since $H_0(x)=1$, each of the prefactors $\phi_0(x_j)$ reduces to
$\exp(-m\omega x_j^2/2\hbar)$. Finally, elementary row and column operations and van der Mond's theorem can be applied as they were in 
\cite{GirWriTri01} to the derivation of the ground state of the harmonically trapped TG gas, to reduce each Slater determinant to a
Bijl-Jastrow product of $(N-1)(N-2)/2$ factors $(x_k-x_\ell)$. The final result is 
\begin{eqnarray}\label{model}
\Psi_{0M}(x_1,\cdots,x_N)&=&\exp(-\sum_{j=1}^N m\omega x_j^2/2\hbar)\nonumber\\
&\times&\sum_{j=1}^N\prod_{1\le k<\ell\le N,(k,\ell)\ne j}(x_k-x_\ell)\ 
\end{eqnarray}
where an irrelevant normalization factor has been dropped.

\textit{Evaluation of} $\Psi_{0B}$ \textit{by duality mapping:} A wave function $\Psi_{0B}$ with Bose symmetry (totally symmetric),
which is an eigenstate of $\hat{H}_0$ with eigenvalue $E_0$ of Eq. (\ref{ground energy}) when all $N$ coordinates are different,
and vanishes when any three are equal, can be found by a duality mapping generalizing that used to find the TG ground state
\cite{Gir60,GirWriTri01}. Each locally antisymmetric factor $(x_k-x_\ell)$ can be converted into an absolute value $|x_k-x_\ell|$
by multiplication by a signum factor $\text{sgn}(x_k-x_\ell)=+1(-1)$, $x_k-x_\ell>0(<0)$, yielding
\begin{eqnarray}\label{ground}
\Psi_{0B}(x_1,\cdots,x_N)&=&\exp(-\sum_{j=1}^N m\omega x_j^2/2\hbar)\nonumber\\
&\times&\sum_{j=1}^N\prod_{1\le k<\ell\le N,(k,\ell)\ne j}|x_k-x_\ell|\ . 
\end{eqnarray}
Since each factor $\text{sgn}(x_k-x_\ell)$ is constant except for a sign change at $x_k=x_\ell$, it follows that except at collision
points $x_k=x_\ell$, $\Psi_{0B}$ is still an eigenstate of $\hat{H}_0$ with eigenvalue $E_0$. It is the desired exact ground state
with three-body hard-core interactions. It is instructive to write it out explicitly for $N=3$ and $N=4$. For $N=3$ Eq. (\ref{ground})
reduces to
\begin{eqnarray}\label{N=3}
&&\Psi_{0B}(x_1,x_2,x_3)=\exp(-\sum_{j=1}^3 m\omega x_j^2/2\hbar)\nonumber\\
&&\times(|x_2-x_3|+|x_1-x_3|+|x_1-x_2)|)\ ,
\end{eqnarray}
which obviously vanishes if $x_1=x_2=x_3$. For $N=4$ one finds
\begin{eqnarray}\label{N=4}
&&\Psi_{0B}(x_1,\cdots,x_4)=\exp(-\sum_{j=1}^4 m\omega x_j^2/2\hbar)\nonumber\\
&&\times(|x_2-x_3||x_2-x_4||x_3-x_4|\nonumber\\
&&+|x_1-x_3||x_1-x_4||x_3-x_4|\nonumber\\
&&+|x_1-x_2||x_1-x_4||x_2-x_4|\nonumber\\
&&+|x_1-x_2||x_1-x_3||x_2-x_3|)\ ,
\end{eqnarray}
which vanishes if $x_2=x_3=x_4$ or $x_1=x_3=x_4$ or $x_1=x_2=x_4$ or $x_1=x_2=x_3$.

\textit{Implicit two-body interactions:} For consistency a three-body hard-core potential term 
$V_3=\lim_{g_3\to +\infty} g_3\sum_{1\le j<k<\ell\le N}\delta(x_j-x_k)\delta(x_j-x_\ell)$ should be added to $\hat{H}_0$ to
exhibit the three-body hard-core interaction. A subtlety that appears to have been missed previously 
is that in 1D, the existence of three-body hard-core interactions automatically generates two-body soft-core interactions
as well. This is not true in 2D, but in 1D it is an unavoidable consequence of the topology of the configuration space, and
holds for any many-boson wave function satisfying the three-body hard-core constraint, whether the exact ground state of
Eq. (\ref{ground}) or an approximate ground state such as the Pfaffian-like ansatz of \cite{ParKeiCir07}, and it holds for
excited states as well. To see this, note that the three-body hard-core constaint region $x_j=x_k=x_\ell$ is an 
$(N-3)$-dimensional hyperline which is
the common intersection of the three $(N-1)$-dimensional hyperplanes $x_j=x_k$, $x_j=x_\ell$, and $x_k=x_\ell$, and these hyperplanes
divide the configuration space into disjoint regions. On the other hand, in 2D the coordinates $z_j$ are complex numbers in the
$(x,y)$ plane, the region $z_j=z_k=z_\ell$ has dimension $2N-6$, and the two-body collision regions each have dimension
$2N-2$, also hyperlines which are easily circumvented, and there is an exact ground state with \emph{only} three-body hard cores,
realized by a state of Pfaffian form \cite{Gre_etal,Rei96}. Returning to the 1D case, suppose, for example, that 
$x_1=x_2<x_4<x_3<\cdots$, and we wish to realize a three-body hard-core interaction at $x_1=x_2=x_3$. Then the particle 
originally at $x_3$ must be moved to the left across the particle at $x_4$, crossing the hyperplane $x_3=x_4$. This 
generates a finite two-body interaction at $x_3=x_4$,  
since local Bose symmetry about this plane requires that the wave function  
$\Psi$ satisfy $\Psi(x_1,x_2,x_3,x_3+\epsilon,x_5,\cdots)=\Psi(x_1,x_2,x_3,x_3-\epsilon,x_5,\cdots)$ for infinitesimal $\epsilon$.
So long as $\partial\Psi(x_1,x_2,x_3,x_3+\epsilon,x_5,\cdots)/\partial\epsilon$ does not vanish, this requires that 
$\partial\Psi(x_1,x_2,x_3,x_4,x_5,\cdots)/\partial x_4$ change sign at $x_4=x_3$, implying that there will be a cusp on the hyperplane $x_3=x_4$. The same proof applies, with the obvious changes of arguments, in the neighborhood of every hyperplane $x_j=x_k$. 
This is the typical behavior for eigenstates of a Hamiltonian containing
a LL two-body interaction term \cite{LieLin63} $V_2=g_2\sum_{1\le j<k\le N}\delta(x_j-x_k)$, where $g_2$ is finite
so long as $\Psi$ does not vanish there, true only for a two-body hard-core interaction (TG limit).  
$g_2$ can be determined in the usual way by integrating Schr\"{o}dinger's equation from $x_j=x_k-$ to $x_j=x_k+$,
or, perhaps more simply, by noting which Dirac delta function terms are generated when the kinetic energy operator acts on
the absolute value factors in $\Psi$, and noting that these must be cancelled by like delta function terms in a two-body
interaction Hamiltonian $V_2$ which must be added to $\hat{H}_0$ to cancel them, in order that $E_0\Psi$ contain no delta functions.
Since the value of $\Psi$
on the hyperplane $x_j=x_k$ depends on all the other coordinates, $g_2$ is not constant as in the usual LL interaction, 
but depends on all the coordinates: $g_2=g_2(x_1,\cdots,x_N)$. For the case $N=3$, one finds after some algebraic reduction
\begin{equation}
g_2(x_1,x_2,x_3)=\frac{\hbar^2}{2m}(|x_1-x_2|+|x_1-x_3|+|x_2-x_3|)^{-1} .
\end{equation}
As $x_3$ approaches the hyperplane $x_1=x_2$, $g_2\to +\infty$, showing that the three-body hard-core
interaction is already implied by the dependence of $g_2$ on $x_3$, and the same argument applies for the other two
hyperplanes. The complexity of the explicit expression increases rapidly with $N$, but $g_2$ is always of the form
$\frac{\hbar^2}{2m}$ times an expression $\frac{\mathfrak{N}}{\mathfrak{D}}$ where the denominator $\mathfrak{D}$ is just
$\Psi_{0B}$ of Eq. (\ref{ground}) with the exponential prefactor omitted, and the numerator $\mathfrak{N}$ is a sum of
products of factors $|x_p-x_q|$ of degree one less than the denominator. It approaches $+\infty$ as any three particles
approach the same point.

\textit{Comparison of exact ground state and Pfaffian-like approximation:} The Pfaffian-like approximate ground state of
Paredes \textit{et al.} \cite{ParKeiCir07} is defined only for even $N$, and is obtained by subdividing the set of $N$ particles into two
subsets of $\frac{N}{2}$ each, assuming that each of these subsets is in its TG ground state, taking the product of these
two $\frac{N}{2}$-particle states, and restoring total Bose symmetry by summing over all ways of selecting
$\frac{N}{2}$ from $N$. They assumed ring geometry, necessitating more complicated ring-periodic TG states, but here the simpler
case of harmonic trapping is assumed. For the simplest case $N=4$ the result is
\begin{eqnarray}\label{PfN=4}
&&\Psi_{0Pf}(x_1,\cdots,x_4)=\exp(-\sum_{j=1}^4 m\omega x_j^2/2\hbar)\nonumber\\
&&\times(|x_1-x_2||x_3-x_4|+|x_1-x_3||x_2-x_4|\nonumber\\
&&+|x_1-x_4||x_2-x_3|)\ ,
\end{eqnarray}
which is to be compared with the exact ground state, Eq. (\ref{N=4}). The Pfaffian-like state is simpler, being built from
products of two factors $|x_p-x_q|$ as compared with products of three in (\ref{PfN=4}), but (\ref{N=4}) is exact whereas
(\ref{PfN=4}) is not an energy eigenstate, although it does vanish if any three particles are at the same point. It is easy
to see that the Pfaffian-like state leads to soft-core LL two-body interactions depending on all four coordinates 
when the kinetic energy Hamiltonian acts
on the absolute values, just as the exact ground state does, in accord with the previous topological proof.  

\textit{Discussion and Outlook:} The exact ground state of a harmonically trapped system of $N\ge 3$ bosons in 1D with three-body
hard-core interactions was found by duality mapping from an ideal gas of mixed symmetry with three-body nodes. It was found to
have two-body soft-core interaction cusps as well, and it was shown that this is an inescapable consequence of the topology
of the configuration space of a system of $N\ge 3$ bosons in 1D with three-body hard-core interactions. In view of previous
suggestions of a possible connection between three-body hard-core interactions and non-Abelian anyons with potential applications
to creation of topologically-protected qubits for quantum computation, it seems worthwhile to look for ways of producing
three-body hard-core interactions in a 1D ultracold Bose gas. This might be facilitated by the fact that the dimensionless
coupling constant for three-body interactions in 1D is a density-independent ratio $\gamma_3=mg_3/\hbar^2$, so higher
densities could be used than are required to reach the two-body hard-core limit $\gamma_2=mg_2/n\hbar^2\gg 1$, which requires
low densities $n$.

\begin{acknowledgments}
Most of this work was carried out while I was participating in the program "Beyond Standard Optical Lattices" 
at the Kavli Institute for Theoretical Physics, University of California, Santa Barbara. I am grateful to the organizers and the KITP Director, David Gross, for the opportunity to participate in this program, to Maciej Lewenstein for a suggestion and discussions which inspired this paper, and to Bel\'{e}n Paredes, Matteo Rizzi, and Marcos Rigol for helpful discussions and suggestions. It was supported at the KITP by the National Science Foundation under grant number NSF PHY05-51164, and at the University of Arizona by the U.S. Army Research Laboratory and the U.S. Army Research Office under grant number W911NF-09-1-0228. 
\end{acknowledgments}
\end{document}